\documentclass[12pt]{article}
\usepackage{amsfonts}
\usepackage{latexsym}
\usepackage{amsmath,amssymb}
\usepackage{verbatim}

\usepackage{setspace}

\usepackage[textheight=9in, textwidth=6.5in, letterpaper]{geometry}
\def\half{{1\over 2}}
\numberwithin{equation}{section}

\def\ip{${\mathcal I}^+$}

\def\cs{{\cal S}}

 \def\p{\partial}
 \def\bz{{\bar z}}
 
\def\0{{(0)}}
\def\1{{(1)}}
\def\2{{(2)}}

\def\ci{{\mathcal I}}

\def\<{\langle }
\def\>{\rangle }
\def\[{\left[}
\def\]{\right]}

\newcommand{\bea}{\begin{eqnarray}}
\newcommand{\eea}{\end{eqnarray}}
\newcommand{\be}{\begin{equation}}
\newcommand{\ee}{\end{equation}}
\newcommand{\ba}{\begin{align}}
\newcommand{\ea}{\end{align}}

\renewcommand{\epsilon}{\varepsilon}

   \makeatletter
  \let\over=\@@over \let\overwithdelims=\@@overwithdelims
  \let\atop=\@@atop \let\atopwithdelims=\@@atopwithdelims
  \let\above=\@@above \let\abovewithdelims=\@@abovewithdelims
\renewcommand\section{\@startsection {section}{1}{\z@}%
                                   {-3.5ex \@plus -1ex \@minus -.2ex}
                                   {2.3ex \@plus.2ex}%
                                   {\normalfont\large\bfseries}}

\renewcommand\subsection{\@startsection{subsection}{2}{\z@}%
                                     {-3.25ex\@plus -1ex \@minus -.2ex}%
                                     {1.5ex \@plus .2ex}%
                                     {\normalfont\bfseries}}

\begin{document}
\begin{titlepage}
\unitlength = 1mm
\ \\
\vskip 4cm
\begin{center}

{ \LARGE {\textsc{Black Hole Information Revisited}}}

\vspace{0.8cm}
Andrew Strominger

\vspace{1cm}

{\it  Center for the Fundamental Laws of Nature, Harvard University,\\
Cambridge, MA 02138, USA}

\begin{abstract} 
We argue that four-dimensional  black hole evaporation inevitably produces an infinite number of soft particles in addition to the thermally distributed `hard' Hawking quanta, and moreover that the soft and hard particles are highly correlated. This raises the possibility  that quantum purity is restored by correlations between the hard and soft radiation, while inclusive measurements which omit the soft radiation observe the thermal Hawking spectrum. In theories whose only stable particle is the graviton, conservation laws are used to  argue that such correlations are in principle sufficient for the soft gravitons to purify the hard thermal ones.   
\end{abstract}

\vspace{1.0cm}

\end{center}

\end{titlepage}

\pagestyle{empty}
\pagestyle{plain}

\def\g{\gamma}
\def\gzz{\gamma_{z\bz}}
\def\gzu{\gamma^{z\bz}}
\def\vx{{\vec x}}
\def\p{\partial}
\def\po{$\cal P_O$}
\def\cN{{\cal N}_\Sigma }
\def\N{${\cal N}_\Sigma ~~$}
\def\G{\Gamma}
\def\l{\ell }
\def\vv{{\vec v}}
\def\hx{\hat x}
\def\hv{\hat v}
\def\hq{\hat q}
\def\D{{\cal D}}
\def\P{{\rm Page}}
\def\ea{${\cal H}_3$}\def\vx{\vec x}
\def\lir{$\Lambda_{IR}$}
\pagenumbering{arabic}

\tableofcontents
\section{Introduction}
According to Hawking \cite{Hawking:1974sw}, quantum black holes emit a nearly thermal spectrum of radiation. This Hawking radiation is `hard' in the sense that the emission probability vanishes as the energy $ \omega \to 0$. From a variety of viewpoints it is known that such a process is impossible in 4D: hard radiated quanta are always accompanied by an infinite cloud of tightly correlated soft quanta. In this note we conjecture that the full evaporation process is unitary, but that the inclusive cross section 
traced over all final soft quanta reproduces the Hawking result at leading order. This proposed unitarity restoration process is an alternative to that of Page \cite{Page:1979tc} in which quantum purity  is restored by correlations between early and late time hard Hawking quanta. We then show that, for the the special case of pure 4D gravity, supertranslation and superrotation conservations laws do imply that tracing over the outgoing soft quanta fully decoheres  the  hard emissions.  Similar conclusions were independently reached in the very recent work of 
Carney {\it et. al.} \cite{Carney:2017jut} who showed, using rather different quantum information-theoretic  methods,  that for  a generic scattering process (not restricted to black holes) tracing over soft quanta greatly and sometimes fully decoheres the final state. The more general case, beyond pure 4D gravity, is left to future work. 

Reversing  the perspective, whether or not soft quanta do fully restore unitary in black hole evaporation, they cannot be ignored. Any process which excludes them - even for a burning lump of coal - will violate unitarity.

This work is a logical outgrowth of the observation \cite{Strominger:2014pwa} that supertranslation charge conservation constrains black hole evaporation,  the analysis of near-horizon soft hair in \cite{hps,Hawking:2016sgy}  and more generally recent insights into symmetries, conservation laws and soft theorems reviewed in \cite{Strominger:2017zoo}. The importance of soft contributions to gravitational entanglement entropy was discussed in \cite{Kapec:2016aqd}. 

\section{Soft particle creation  by black holes}
Consider a black hole which is formed by gravitational collapse and then eventually evaporates and disappears. According to Hawking's semiclassical computation, the outgoing Hawking radiation has a nearly thermal (up to greybody factors and sub-semiclassical corrections) blackbody spectrum, in which the probability per unit time of emitting a boson of energy $\omega$ is approximately 
\be \label{no} N(\omega)\sim {\omega^2 \over e^{ \omega /T_H}-1}~.\ee
Here the Hawking temperature for a black hole of mass $M$ is
\be T_H={1 \over 8\pi GM}.\ee
After the black hole has evaporated, Hawking argued \cite{Hawking:1976ra}, 
the final outgoing state is a density matrix
\be\label{mixd} |\Psi_{\rm in} \> \to \rho_{\rm Hawking}=\sum_\alpha\rho_\alpha |H_\alpha\>\< H_\alpha |. \ee 
Here the outgoing states $|H_\alpha\>$ have particle occupation numbers roughly of the form \eqref{no} and the index $\alpha$ runs over all the states in the thermal radiation ensemble.  Since \eqref{mixd} entails a breakdown of determinism,  this conclusion is hard to swallow. 

One might expect instead \cite{Page:1979tc} that there   is some yet-to-be-found flaw in the reasoning of  \cite{Hawking:1976ra} and that there is a unitary process of the form 
\be\label{unti}  \cs |\Psi_{\rm in} \> = \sum_\alpha c_\alpha |H_\alpha\>~, \ee
for some constants $c_\alpha$ which we do not yet understand how to compute. 
In fact, the literal form \eqref{unti} is generically impossible for a simple reason. Picture  the  black hole as some kind of unitary black box which spews out quanta from the horizon in a pure state which is some  element in the ensemble \eqref{mixd}. These quanta will have IR divergent exchanges of soft gravitons which occur far outside the horizon and are insensitive to the details of the black box. Exponentiating these exchanges will give $e^{-\infty}$ and set the amplitude to zero, just as soft photon exchanges  do for $e^+e^-\to e^+e^-$ in QED. The amplitude to produce any state with quanta distributed as in \eqref{no} will vanish. Nonvanishing amplitudes must have  poles in the number of soft quanta as $\omega\to 0$. This conclusion can alternately be viewed as a consequence of supertranslation charge conservation \cite{Strominger:2013jfa,Kapec:2017tkm} as we review below.

Put another way, due to IR divergences, Hawking's leading order semiclassical result for unaccompanied hard radiation is not close to the exact answer.
It  is corrected by an infinite amount.  

 This type of IR problem is familiar in quantum field theory. The usual fix is to compute inclusive cross sections with an IR cutoff. This is fine if we are computing collider cross sections since real  detectors have an intrinsic IR cutoff. But subtle issues of unitarity cannot be addressed in the context of inclusive cross sections. For that we need an $\cs$-matrix. 

In QED a second type of  fix for this type of IR problem, which does  involve an $\cs$-matrix, was given long ago by 
Faddeev and Kulish (FK) \cite{Kulish:1970ut}. The FK construction was recently extended to gravity in \cite{Ware:2013zja}.  All asymptotic hard particles are 
dressed with a cloud of soft gravitons involving $\omega\to 0$ poles.  The IR hard-particle pair exchange divergences are cancelled by divergences in exchanges between hard particles and soft graviton clouds. $\cs$-matrix elements of the dressed hard particles are IR finite. 

There is a large ambiguity in the FK choice of the dressing. IR finiteness only fixes the coefficients of the $\omega\to 0$ poles in the dressing factors. For the purposes of IR safety, one can take the dressing to be entirely composed of gravitons below an arbitrarily low energy scale \lir.

This suggests a natural modification of \eqref{unti}. We propose it be modified to  
\be\label{nti}  \cs |\Psi_{\rm in} \> = \sum_\alpha c_\alpha |H_\alpha\>|S_\alpha\>=|\Psi_{\rm out} \>  , \ee
where we have decomposed  the Hilbert space into hard and soft factors, which are separated by a scale \lir$\ll T_H$. The Hawking radiation $ |H_\alpha\>$ is in the hard part and the soft part $ |S_\alpha\>$ contains the dressing with the soft poles required for IR finiteness. 
This simply states that thermally-distributed hard particles are accompanied by a non-thermal 
cloud of soft particles. While we know that {\it some} soft dressing must be present, we do not know how to compute the exact soft spectrum. 

We now conjecture  that the soft factors associated to every distinct hard factor are orthogonal:
\be\label{nm} \<S_\alpha|S_\beta\>=\delta_{\alpha\beta}.\ee
Physically this means that there is a unique soft emission spectrum for every hard state. Since there is an infinite degeneracy of soft states this is not apriori ruled out. We will make plausibility arguments  for \eqref{nm} in a restricted context in the next section but for now we just regard \eqref{nm} as a conjecture. 
It follows from \eqref{nm} that detectors which only observe the hard Hawking radiation, and miss the soft dressings, will observe a density matrix
\be \rho_H =\sum_\alpha c_\alpha  c_\alpha ^*|H_\alpha \>\< H_\alpha  |.\ee
Taking 
\be \label{eg} c_\alpha =\sqrt{\rho_\alpha }e^{i\theta_\alpha }, ~~~~  |\Psi_{\rm out} \> = \sum_\alpha \sqrt{\rho_\alpha }e^{i\theta_\alpha } |H_\alpha\>|S_\alpha\>\ee
where $\rho_\alpha $ is defined  in \eqref{mixd}, we find 
\be \rho_H=\rho_{\rm Hawking}.\ee
This motivates the conjecture that black hole evaporation is a unitary process
of the form \eqref{nti} with the magnitude of $c_\alpha $ given by \eqref{eg} at leading semiclassical order. A detector which cannot capture  the soft radiation will see  the Hawking spectrum.\footnote{Confirming  unitarity would require at a minimum  measuring the  soft pole using  gravitational memory detectors \cite{Strominger:2014pwa}.} In particular there will be no correlations between Hawking quanta 
before and after the Page time as put forth in \cite{Page:1979tc}. Rather the information is stored in correlations between the hard and soft quanta.\footnote{ As it was recently shown in \cite{Carney:2017jut} that soft and hard quanta are generically highly entangled in any  scattering process, even for a  burning lump of coal some of the information is contained in such correlations. }

The Page proposal \eqref{unti} did not provide any algorithm for how the constants 
$c_\alpha$ determining the out states are computed, and in the intervening decades no such algorithm  has clearly emerged. In contrast in the present proposal  the magnitude of $c_\alpha$ in \eqref{nti} is given at leading order by the Hawking computation as $\sqrt{\rho_\alpha}$, but no algorithm is provided here for determining the relative phases $\theta_\alpha $ of the outgoings states. That would require insights and methods beyond those of the present work.

\section{A restricted example}

In the previous section we made a rather general proposal that thermal Hawking radiation is purified by correlations with soft quanta. In this section we put flesh on this proposal and show that the key conjecture \eqref{nm} holds for a simplified 4D setting in which 
gravitons are the only stable  particles. In this case, there should be a unitary $\cs$-matrix with  purely gravitational in and out states. We do not know of a concrete 4D example of such a theory, but IIB string theory is a  10D  example in which the only stable particles are the  graviton and its superpartners.  In the last  section
the generalization to theories with multiple stable massive or massless particles is briefly discussed.

We seek more detail about the soft clouds  $|S_\alpha \>$ arising from a given $|H_\alpha \>$. 
The crucial pole structure can be determined from supertranslation charge conservation \cite{Strominger:2013jfa} following a similar analysis given for QED in \cite{Kapec:2017tkm}.  There is one conserved supertranslation charge for every angle $(z,\bz)$ on the celestial sphere at null infinity. In a convenient basis the outgoing charge can be decomposed into soft and hard pieces as 
\bea Q^+(z,\bz)&=&Q^+_S(z,\bz)+Q^+_H(z,\bz)-{M\gamma_{z\bz}\over 4\pi}~,\cr
Q^+_H&=&{\gamma_{z\bz} \over 4\pi G}\int_{\ci^+} du T_{uu}~,\cr
Q^+_S&=&-{\gamma_{z\bz}\over 16\pi G} \int_{\ci^+} du \left(D_z^2N^{zz}+D_\bz^2N^{\bz\bz}\right)~,
\eea
where in the first line we have subtracted off the zero mode proportional to the total mass which is not our interest here. 
In the conventions of \cite{Strominger:2017zoo} $u$ is retarded time,  $D_z$ is the covariant derivative on the unit sphere with metric $\gamma_{z\bz}$, $T_{uu}(u,z,\bz)$  (and $T_{uz}$ below) is a    coefficient of the 
$r^{-2}$ term in the expansion about \ip\ of the  gravity wave stress tensor 
and 
\be N_{zz}(u,z,\bz)=\p_uC_{zz}(u,z,\bz) \ee is the Bondi news with $C_{zz}$ the leading transverse perturbation  of the asymptotically  flat metric.  $Q^+_H$ is the hard energy flux through the angle $(z,\bz)$. $Q^+_S$ is a soft graviton zero mode at the same angle. Supertranslation charge conservation equates the total $Q^+$ to  a similar expression for the incoming charge $Q^-$  on $\ci^-$. However for simplicity we restrict to black holes especially prepared in a `neutral' eigenstate with\footnote{Asymptotic FK dressed particles states are constructed to commute with the  supertranslation charges and hence convenient for making such a neutral black hole.} 
\be\label{dro} Q^-|\Psi_{\rm in} \> =0.\ee It follows that $Q^+|\Psi_{\rm out} \>=0$ or
\be \label{rtu} Q^+_H(z,\bz)|\Psi_{\rm out} \> =-Q^+_S(z,\bz)|\Psi_{\rm out} \> . \ee
The hard part of the outgoing state $|H_\alpha  \>$ is a collection of hard gravitons. 
Ignoring, for the rest of this paragraph, the spin of the graviton, such  states  can be  characterized by the positions $z_\alpha ^k$ where the outgoing
Hawking quanta exit through \ip\ with energies $E^k_\alpha $.  Such momentum eigenstates diagonalize the hard supertranslation charge 
\be\label{sxz}  Q_H^+(z,\bz)|H_\alpha  \> =\sum_k {E_\alpha ^k \delta^2(z-z_\alpha ^k})  |H_\alpha  \> .\ee
Since black holes tend to emit at low angular momenta, it might seem more natural to use an angular momentum basis. However such bases do not diagonalize $Q^+_H$ and therefore do not obey a simple relation of the form \eqref{sxz}. 
It then follows from \eqref{rtu} that the soft modes must obey 
\be  Q_S^+(z,\bz)|S_\alpha  \> =-\sum_k {E_\alpha ^k \delta^2(z-z_\alpha ^k} )|S_\alpha  \> .\ee
Soft clouds  which obey such relations were constructed (for QED)  in \cite{Kapec:2017tkm}, and necessarily produce $\omega\to 0$ poles in field expectation values. Up to the spin, the eigenvalue in \eqref{sxz}  almost uniquely determines the hard state. A counterexample is the case of  two gravitons at {\it exactly} the same $z_k$ each with energy $\half E_k$. This is not an issue in the present  context because the probability of producing two {\it exactly} collinear gravitons is zero in the Hawking formula.   Since eigenstates of a hermitian operator with distinct eigenvalues are orthogonal we may then normalize so that the soft states $|S_\alpha \>$  obey \eqref{nm}.

Now let us account for  spin by adding an extra label $s^k_\alpha =\pm 2$ to each hard graviton.  Two states $|H_\alpha \>$ which differ only in the spins of one or more gravitons will have the same eigenvalue of $Q^+_H$ in \eqref{sxz}. However the soft state is further constrained by conservation of superrotation charge $Q^\pm_z$ at each angle on the celestial sphere \cite{Cachazo:2014fwa,Kapec:2014opa}. We now show that these charges do discern the spin and force the soft clouds to be orthogonal if the hard graviton spins are not aligned.\footnote{This was  suggested in \cite{Carney:2017jut}.} As in \eqref{dro} we restrict to 
\be\label{dros} Q^-|\Psi_{\rm in} \> =0=Q^+|\Psi_{\rm out} \>.\ee
The outgoing superrotation charge can be decomposed into soft and hard pieces (in the conventions of \cite{Strominger:2017zoo}) as 
\bea Q^+_z(z,\bz)&=&Q^+_{Sz}(z,\bz)+Q^+_{Hz}(z,\bz)~,\cr
Q^+_{Hz}&=&{1 \over 8\pi G} \int_{\ci^+} du (T_{uz}-uD_zT_{uu})~,\cr
Q^+_{Sz}&=&-{1 \over 16\pi G}  \int_{\ci^+} du D_z^3C^{zz}~.
\eea
$Q^+_{Hz}$ is essentially the  `twist'  (the hard angular momentum  plus  boost weight $E\p_E$ ) radiated through  $(z,\bz)$:\footnote{Here we have taken $Y^z=\gamma^{z\bz}\delta^2(z-z^k_\alpha),~Y^\bz=0$ in equation (5.4) of  \cite{Kapec:2014opa} and integrated by parts.} 
\be\label{sxdz}  Q_{Hz}^+|H_\alpha  \> = {i \over 2} \sum_k D^\bz \delta^2(z-z_\alpha ^k)  ( s_\alpha ^k-2 -E_\alpha ^k\p_{E_\alpha ^k}) |H_\alpha  \> .\ee
Since our interest here is in determining the effect of the spin on $|S_m\>$, we construct a conserved charge involving only the spin by integrating an appropriate linear combination of $Q_{Hz}^+$ 
and $Q_{H\bz}^+$ over a small region $R^k_\alpha$ surrounding any point $z^k_\alpha$.
Let us define 
\be Q^+_H( z^k_\alpha)=-i\int_{R^k_\alpha}d^2z\gamma_{z\bz}\big((z-z^k_\alpha)Q_{Hz}^+-(\bz-\bz^k_\alpha)Q_{H\bz}^+\big)~,\ee
with a similar expression for $Q^+_S( z^k_\alpha)$.
We then have 
\be Q^+_H( z^k_\alpha)|H_\alpha\>=s^k_\alpha|H_\alpha\>~,\ee
and from \eqref{dros}
\be Q^+_S( z^k_\alpha)|S_\alpha\>=-s^k_\alpha|S_\alpha\>~ .\ee
This constrains the behavior of the soft cloud at subleading order in the soft expansion,\footnote{Note that there is no IR divergence associated to superrotations. Constraints of this type on the soft state might generalize the present paper above $D=4$.}  and implies that all (noncollinear) hard graviton states appearing in the evaporation of an initially uncharged black hole are accompanied by a different soft state. 
This is then enough information to conclude that the soft graviton clouds obey \eqref{nm}.

\section{Comments}

More massless particles, beyond the graviton,  produces degeneracies in the supertranslation and superrotation charges, which can no longer on their own distinguish all states in the Hawking ensemble. At the same time they also introduce new infinities of conserved charges, possibly even in the case of scalars \cite{Campiglia:2017dpg}, which may further correlate the hard and soft factors. Moreover we needed here only half (rotations, and not boosts) of the correlations imposed by superrotation charge conservation, and there are perhaps further unexplored correlations from conservation laws associated to the sub-subleading soft theorem \cite{Cachazo:2014fwa,Campiglia:2016jdj,Campiglia:2016efb,Conde:2016rom,Laddha:2017ygw}.
In the real world there are stable massive particles such as the electron. Their effects on the present analysis might best proceed via the hyperbolic resolution of future timelike infinity
initiated in \cite{Campiglia:2015qka}, which brings in concepts from AdS$_3$/CFT$_2$. 
It is certainly possible however that counterexamples to the central conjecture \eqref{nm} can be found beyond the simple example of the preceding section.\footnote{Certainly theories with exact global symmetries will violate \eqref{nm}, since two different hard states related by the symmetry will produce the same soft state. However it is believed such global symmetries  cannot arise  in quantum gravity.}  If so, one might impose \eqref{nm} as a consistency condition restricting sensible quantum theories of gravity. 
Another open issue is the extension to higher dimensions.  Soft theorems and their associated  conservation laws still exist and correlate the soft state accompanying any hard emission, but the structure of IR divergences is very different.

The analysis of section 3 makes a number of simplifying restrictions including to  incoming states with no  supertranslation or superrotation charges. Although highly non-generic, the information paradox is still present and can be addressed with these restrictions. It would be interesting  to extend the analysis to generic incoming states. 

We have outlined an alternative  pattern for information return 
which relies on non-thermal soft graviton production in the black hole evaporation process. Observers equipped with only hard particle detectors see the mixed thermal radiation state predicted by Hawking.  Corrections to the Hawking calculation arise only  in the regime in which we know it must break down due to IR divergences. We do not have a proposal here for a complete calculation of $|\Psi_{\rm out} \>$ including soft modes, which would  include a determination of  the phases in  \eqref{eg}.

We leave these issues to future investigation.

\section*{Acknowledgements}
I am   grateful  to T. Dumitrescu, S. Giddings, D. Harlow, S. Hawking, D. Kapec, P. Mitra, S. Pasterski, M. Perry, A. Raclariu, G. Semenoff and  S. Zhiboedov for useful conversations. This work was supported by NSF  grant  1205550 and the John Templeton Foundation.

\end{document}